\documentclass[twocolumn]{IEEEtran}

\usepackage[T1]{fontenc}
\usepackage{amsmath}
\makeatletter
\AtBeginDocument{\let\comt\@firstofone}
\makeatother

\usepackage[cmintegrals]{newtxmath}
\usepackage[nolist]{acronym}
\usepackage{color,soul}
\usepackage{graphicx}
\usepackage{setspace}
\usepackage{soul}

\usepackage{psfrag}
\newcommand{\comt}[1]{\textcolor{blue}{#1}}

\usepackage{comment}
\begin{document}

\title{Boosting 5G on Smart Grid Communication: \\ A Smart RAN Slicing Approach}
%
%

\author{Dick~Carrillo,
        Charalampos~Kalalas,
        Petra~Raussi,
        Diomidis S.~Michalopoulos,
        Demóstenes Z. Rodríguez,\\
        Heli Kokkoniemi-Tarkkanen,
        Kimmo Ahola,
         Pedro H. J. Nardelli,
         Gustavo Fraidenraich, and
         Petar Popovski
}

\markboth{ }%
{Shell \MakeLowercase{\textit{et al.}}: Bare Demo of IEEEtran.cls for IEEE Communications Society Journals}
\maketitle

\begin{abstract}
\comt{Fifth-generation (5G) and beyond systems are expected to accelerate the ongoing transformation of power systems towards the smart grid.
However, the inherent heterogeneity in smart grid services and requirements pose significant challenges towards the definition of a unified network architecture.
In this context, radio access network (RAN) slicing emerges as a key 5G enabler to ensure interoperable connectivity and service management in the smart grid.
This article introduces a novel RAN slicing framework which leverages the potential of artificial intelligence (AI) to support IEC 61850 smart grid services.
With the aid of deep reinforcement learning, efficient radio resource management for RAN slices is attained, while conforming to the stringent performance requirements of a smart grid self-healing use case.
Our research outcomes advocate the adoption of emerging AI-native approaches for RAN slicing in beyond-5G systems, and lay the foundations for differentiated service provisioning in the smart grid.}
\end{abstract}


\begin{acronym}
  \acro{1G}{first generation of mobile network}
  \acro{1PPS}{1 pulse per second}
  \acro{2G}{second-generation of mobile network}
  \acro{3G}{third-generation of mobile network}
  \acro{4G}{fourth-generation of mobile network}
  \acro{5G}{fifth-generation}
  \acro{ARQ}{automatic repeat request}
  \acro{ASIP}{application specific integrated processors}
  \acro{AWGN}{additive white Gaussian noise}
   \acro{BER}{bit error rate}
  \acro{BCH}{Bose-Chaudhuri-Hocquenghem}
  \acro{BRICS}{Brazil-Russia-India-China-South Africa}
  \acro{BS}{base station}
  \acro{CDF}{cumulative density function}
  \acro{CoMP} {cooperative multi-point}
  \acro{CP}{cyclic prefix}
  \acro{CR}{cognitive radio}
  \acro{CS}{cyclic suffix}
  \acro{CSI}{channel state information}
  \acro{CSMA}{carrier sense multiple access}
  \acro{DFT}{discrete Fourier transform}
  \acro{DFT-s-OFDM}{DFT spread OFDM}
  \acro{DRL}{deep reinforcement learning}
  \acro{DSA}{dynamic spectrum access}
  \acro{DVB}{digital video broadcast}
  \acro{DZT}{discrete Zak transform}
  \acro{eMBB}{enhanced mobile broadband}
  \acro{EPC}{evolved packet core}
  \acro{FBMC}{filterbank multicarrier}
  \acro{FDE}{frequency-domain equalization}
  \acro{FDMA}{frequency division multiple access}
  \acro{FD-OQAM-GFDM}{frequency-domain OQAM-GFDM}
  \acro{FEC}{forward error control}
  \acro{F-OFDM}{Filtered Orthogonal Frequency Division Multiplexing}
  \acro{FPGA}{field programmable gate array}
  \acro{FTN}{faster than Nyquist}
  \acro{FT}{Fourier transform}
  \acro{FSC}{frequency-selective channel}
  \acro{GFDM}{generalized frequency division multiplexing}
  \acro{GPS}{global positioning system}
  \acro{GS-GFDM}{guard-symbol GFDM}
  \acro{IARA}{Internet Access for Remote Areas}
  \acro{ICI}{intercarrier interference}
  \acro{ICT}{information and communication technology}
  \acro{IDFT}{Inverse Discrete Fourier Transform}
  
  \acro{IFI}{inter-frame interference}
  \acro{i.i.d.}{independent and identically distributed}
  \acro{IMS}{IP multimedia subsystem}
  \acro{IoT}{internet of things}
  \acro{IP}{Internet Protocol}
  \acro{ISI}{intersymbol interference}
  \acro{IUI}{inter-user interference}
  \acro{LDPC}{low-density parity check}
  \acro{LLR}{log-likelihood ratio}
  \acro{LMMSE}{linear minimum mean square error}
  \acro{LTE}{long-term evolution}
  \acro{LTE-A}{Long-Term Evolution - Advanced}
  \acro{M2M}{Machine-to-Machine}
  \acro{MA}{multiple access}
  \acro{MAR}{mobile autonomous reporting}
  \acro{MEC}{mobile edge computing}
  \acro{MF}{Matched filter}
  \acro{MIMO}{multiple-input multiple-output}
  \acro{MMSE}{minimum mean squared error}
  \acro{MRC}{maximum ratio combiner}
  \acro{MSE}{mean-squared error}
  \acro{MTC}{machine-type communication}
  \acro{NEF}{noise enhancement factor}
  \acro{NFV}{network functions virtualization}
  \acro{OFDM}{orthogonal frequency division multiplexing}
  \acro{OOB}{out-of-band}
  \acro{OOBE}{out-of-band emission}
  \acro{OQAM}{offset quadrature amplitude modulation}
  \acro{PAPR}{peak-to-average power ratio}
  \acro{PDF}{probability density function}
  \acro{PHY}{physical}
  \acro{QAM}{quadrature amplitude modulation}
  \acro{PSD}{power spectrum density}
  \acro{QoE}{quality of experience}
  \acro{QoS}{quality of service}
  \acro{RC}{raised cosine}
  \acro{RRC}{root raised cosine}
  \acro{RTT} {round trip time}  
  \acro{SC}{single carrier}
  \acro{SC-FDE}{Single Carrier Frequency Domain Equalization}
  \acro{SC-FDMA}{Single Carrier Frequency Domain Multiple Access}
  \acro{SDN}{software-defined network}
  \acro{SDR}{software-defined radio}
  \acro{SDW}{software-defined waveform}
  \acro{SEP}{symbol error probability}
  \acro{SER}{symbol error rate}
  \acro{SIC}{successive interference cancellation}
  \acro{SINR}{signal-to-interference-and-noise ratio }
  \acro{SMS}{Short Message Service}
  \acro{SNR}{signal-to-noise ratio}
  \acro{STC}{space time code}
  \acro{STFT}{short-time Fourier transform}
  \acro{TD-OQAM-GFDM}{time-domain OQAM-GFDM}
  \acro{TTI}{time transmission interval}
  \acro{TR-STC}{Time-Reverse Space Time Coding}
  \acro{TR-STC-GFDMA}{TR-STC Generalized Frequency Division Multiple Access}
  \acro{TVC}{ime-variant channel}
  \acro{UFMC}{universal filtered multi-carrier}
  \acro{UF-OFDM}{Universal Filtered Orthogonal Frequency Multiplexing}
  \acro{UHF}{ultra high frequency}
  \acro{URLL}{Ultra Reliable Low Latency}
  \acro{V2V}{vehicle-to-vehicle}
  \acro{V-OFDM}{Vector OFDM}
  \acro{ZF}{zero-forcing}
  \acro{ZMCSC}{zero-mean circular symmetric complex Gaussian}
  \acro{W-GFDM}{windowed GFDM}
  \acro{WHT}{Walsh-Hadamard Transform}
  \acro{WLAN}{wireless Local Area Network}
  \acro{WLE}{widely linear equalizer}
  \acro{WLP}{wide linear processing}
  \acro{WRAN}{Wireless Regional Area Network}
  \acro{WSN}{wireless sensor networks}
  \acro{ROI}{return on investment}
  \acro{NR}{new radio}
  \acro{SAE}{system architecture evolution}
  \acro{E-UTRAN}{evolved UTRAN}
  \acro{3GPP}{third generation partnership project}
  \acro{MME}{mobility management entity}
  \acro{S-GW}{serving gateway}
  \acro{P-GW}{packet-data network gateway}
  \acro{eNodeB}{evolved NodeB}
  \acro{UE}{user equipment}
  \acro{DL}{downlink}
  \acro{UL}{uplink}
  \acro{LSM}{link-to-system mapping}
  \acro{PDSCH}{physical downlink shared channel}
  \acro{TB}{transport block}
  \acro{MCS}{modulation code scheme}
  \acro{ECR}{effective code rate}
  \acro{BLER}{block error rate}
  \acro{CCI}{co-channel interference}
  \acro{OFDMA}{orthogonal frequency-division multiple access}
  \acro{LOS}{line-of-sight}
  \acro{VHF}{very high frequency}
  \acro{pdf}{probability density function}
  \acro{ns-3}{Network simulator 3}
  \acro{Mbps}{mega bits per second}
  \acro{EH}{energy harvesting}
  \acro{SWIPT}{simultaneous wireless information and power transfer}
  \acro{AF}{amplify-and-forward}
  \acro{DF}{decode-and-forward}
  \acro{WIT}{wireless information transfer}
  \acro{WPT}{wireless power transfer}
  \acro{FSFC}{frequency selective fading channel}
  \acro{DC}{direct current}
  \acro{FFT}{fast Fourier transform}
  \acro{RF}{radio frequency}
  \acro{SISO}{single-input single-output}
  \acro{RRC}{root raised cosine}
  \acro{TSR}{time-switching relaying}
  \acro{IFFT}{inverse fast Fourier transform}
  \acro{LIS}{large intelligent surfaces}
  \acro{uRLLC}{ultra-reliable low-latency communication}
  \acro{ZMCSCG}{zero mean circularly symmetric complex Gaussian}
  \acro{PPSINR}{post-processing SINR}
  \acro{mMTC}{massive machine-type communication}
  \acro{NR}{New radio}
  \acro{RIS}{reconfigurable intelligent surface} 
  \acro{RAN}{radio access network}
  \acro{i.i.d.}{independent and identically distributed}
  \acro{NOMA}{non-orthogonal multiple access}
  \acro{D2D}{device-to-device}
  \acro{LPWAN}{low-power wide-area network}
  \acro{WiMAX}{worldwide interoperability for microwave access}
  \acro{KPI}{key performance indicator}
  \acro{PLC}{power line communication}
  \acro{PER}{packet-error rate}
  \acro{E2E}{end-to-end}
  \acro{eMTC}{evolved machine-type communication}
  \acro{ISM}{industrial, scientific, and medical}
  \acro{QoS}{quality-of-service}
  \acro{ETSI}{European Telecommunications Standards Institute}
  \acro{P/S-GW}{packet/service gateway}
  \acro{PCRF}{policy and charging rules function}
  \acro{HSS}{home subscriber server}
  \acro{BBU}{baseband unit}
  \acro{NG-RAN}{next-generation RAN}
  \acro{gNB}{next-generation NodeB}
  \acro{CU}{centralized unit}
  \acro{DU}{distributed unit}
  \acro{RU}{radio unit}
  \acro{NSSMF}{network slice subnet management function}
  \acro{NFMF}{network function management function}
  \acro{NSST}{network slice subnet template}
  \acro{VIM}{virtualized infrastructure management}
  \acro{VNFM}{virtualized network function management}
  \acro{VNF}{virtualized network function}
  \acro{VM}{virtual machine}
  \acro{MANO}{ETSI management and orchestration}
  \acro{AI}{artificial intelligence}
  \acro{SLAs}{service-level agreements}
  \acro{NSSI}{network slice subnet instance}
  \acro{E2E-NSI}{end-to-end network slice instance}
  \acro{NSI}{network slice instance}
  \acro{S-NSSAI}{single network slice selection assitance information}
  \acro{NSSAI}{network slice selection assistance information}
  \acro{SST}{slice/service type}
  \acro{SD}{slice differentiator}
  \acro{MIoT}{massive Internet of things}
  \acro{V2X}{vehicle-to-everything}
  \acro{IEC}{International Electrotechnical Commision}
  \acro{LN}{logical node}
  \acro{IAB}{integrated access and backhaul}
  \acro{SG}{smart grid}
  \acro{PV}{Photo-voltaics}
  \acro{DER}{distributed energy resource}
  \acro{ICT}{information and communication technology}
  \acro{AC}{alternating current}
  \acro{SSCI}{subsynchronous control interaction}
  \acro{SSO}{subsynchronous oscillation}
  \acro{WEC}{wind energy converter}
  \acro{VCS}{voltage control system}
  \acro{HV}{high voltage}
  \acro{MV}{medium voltage}
  \acro{LV}{low voltage}
  \acro{DA}{distribution automation}
  \acro{DG}{distribured generation}
  \acro{HVAC}{high voltage AC}
  \acro{HVDC}{high voltage DC}
  \acro{GOOSE}{generic object oriented substation event}
  \acro{SMV}{sampled measured value}
  \acro{TSO}{transmission system operator}
  \acro{DSO}{distribution system operator}
  \acro{SCADA}{supervisory control and data acquisition}
  \acro{EMS}{energy management system}
  \acro{EV}{electric vehicle}
  \acro{ST}{smart transformer}
  \acro{SDEN}{software defined energy network}
  \acro{E2E}{end-to-end}
  \acro{ROI}{return on investment}
  \acro{PMU}{phasor measurement unit}
  \acro{DR}{demand response}
  \acro{EMC}{electromagnetic compatibility}
  \acro{DG}{distributed generation}
  \acro{GSSE}{generic substation status event}
  \acro{SV}{sampled value}
  \acro{MMS}{manufacturing message specification}
  \acro{IED}{intelligent electronic device}
  \acro{MEC}{mobile edge computing}
  \acro{RANA}{RAN architecture}
  \acro{RIL}{RAN isolation level}
  \acro{RSF}{RAN slicing function}
  \acro{UNFS}{upper network functions selection}
  \acro{MAC}{media access control}
  \acro{GA}{grant-free access}
  \acro{AMI}{advanced metering infrastructure}
  \acro{CPU}{central processing unit}
  \acro{RAT}{radio access technologie}
  \acro{RSM}{RAN slicing management}
  \acro{CNFB}{core network functions bridge}
  \acro{NB-IoT}{narrowband IoT}
  \acro{MMSE-SIC}{MMSE-successive interference cancellation}
  \acro{NPN}{non-public networks}
  \acro{MANO}{management and orchestration}
  \acro{NFVO}{NFV orchestrator}
  \acro{NFV-MANO}{NFV management and orchestration}
  \acro{SLA}{service-level agreement}
  \acro{gNodeB}{5G Node B}
  \acro{SSC}{smart substation controller}
  \acro{MU}{merging unit}
  \acro{IRC}{interference rejection combining}
  \acro{ICIC}{inter-cell interference coordination}
  \acro{IRSS}{Intelligent RAN slicing scheduler}
  \acro{RL}{reinforcement learning}
  \acro{MNO}{mobile network operator}
  \acro{RB}{resource block}
  \acro{CBC}{circuit breaker control}
\end{acronym}

\IEEEpeerreviewmaketitle

\vspace{-.4cm}
\section{Introduction and Motivation}
\IEEEPARstart{I}{n} recent years, the ongoing modernization of the aging power systems towards the smart grid has mainly relied on three prevailing trends: (i) \textit{large-scale information acquisition}, with the massive deployment of smart meters to keep track of energy consumption; (ii) \textit{reliable monitoring, protection, and control}, where \acp{IED} offer situational awareness and rapid fault detection; and (iii) \textit{integration of distributed energy resources (DERs)}, which results in high power system dynamics and a growing need for real-time grid supervision to ensure stability. On top of the drivers mentioned above, the deregulation of energy markets and the need for advanced security against hostile cyberattacks have a significant impact on the transformation of power systems.

Instrumental to this paradigm shift is the underlying communication infrastructure deployed for robust, scalable, and reliable connectivity among the power system components \cite{GUN11}.
Connectivity in power systems currently involves a \comt{plethora} of technologies, ranging from optical fibers and power line communication to wireless technologies and satellite networks.
Among those, wired connectivity \comt{schemes} have been extensively used in power systems for localized mission-critical applications.
Notwithstanding, wireless solutions exploit the advantages of lower deployment/maintenance cost and their intrinsic scalable characteristics to offer enhanced grid functionalities \cite{HO13}.
For example, in distribution automation, \acp{IED} need to timely exchange protection-related messages for fast decision-making to avoid extensive disturbances to the entire \comt{grid}. On the other hand, advanced metering installations require highly scalable network deployments to \comt{manage} meter readings from consumers located at disparate spatial locations.
The advent of \ac{5G} and beyond communication networks is expected to revolutionize traditional power systems by supporting a wide range of real-time and autonomous operations \cite{8067687}. 
Unlike previous mobile network generations, \ac{5G} systems are designed to enable three \comt{key} generic services with broadly diverging operational requirements, i.e., \ac{eMBB}, \ac{mMTC}, and \ac{uRLLC}. 
Cellular networks are \comt{progressively} becoming ubiquitous with \comt{omnipresent applicability in} vertical industries.
Thus, the business potential of \ac{5G} in the smart grid domain can be \comt{remarkably} high with the realization of \comt{unprecedented} use cases\comt{, such as millisecond-level precise load control, decentralized fault detection and self-healing operation, and predictive maintenance of grid infrastructure} \cite{5G_microperators_ahokangas}. 
Among the \comt{pivotal 5G} novelties,
\ac{RAN} slicing \comt{allows the partition} of radio resources into logically 
isolated radio networks, \comt{each one} interpreted as a RAN slice.
\ac{RAN} slicing \comt{is recently gaining momentum} as an enabling platform for the integration of \comt{vertical services} over a shared \comt{physical infrastructure}. 

\begin{table*}[t!]
\caption{Categorization of 5G-enabled smart grid services and their relationship with network slices}
\vspace{-.2cm}
\centering
\small{
\begin{tabular}{ |p{4.2cm}|p{1.7cm}|p{1.7cm}|p{1.7cm}|p{1.76cm}|p{2.15cm} |p{1.6cm} |}
\hline
   \textbf{Service category} &\textbf{Latency}&  \textbf{Reliability} & \textbf{Bandwidth}  &  \textbf{Node density} &  \textbf{Service priority} & \textbf{SST value}\\
\hline\hline
 \textbf{Smart distribution automation} &
 Low &
 High &
 Low &
 Medium/Low &
 High &
 2~(uRLLC)\\
\hline
 \textbf{Wide-area monitoring, control,\newline and protection} &
 Low/Medium &
 High/Medium &
 Medium/Low &
 Medium &
 Medium/High &
 1\,\,(eMBB), 2~(uRLLC), 3~(mMTC)\\
\hline
\textbf{Metering data acquisition} & 
 Medium/High &
 Medium/Low &
 Medium/High &
 High &
 Low/Medium &
 3~(mMTC)\\
\hline
 \textbf{Distributed generation \newline integration and microgrids} & 
 Low/Medium &
 High &
 Low &
 Medium/High &
 Medium/High &
 2~(uRLLC), 3~(mMTC)\\
\hline
 \textbf{Volume and price balancing} & 
 Medium/High &
 Medium/Low &
 Medium/High &
 Medium/High &
Low &
1\,\,(eMBB), 3~(mMTC)\\
\hline
\end{tabular}
}
\vspace{-.41cm}
\label{table:requirements_servicescenarios}
\end{table*}

\comt{One of the key challenges for the realization of \ac{RAN} slicing relates to the efficient resource management among \ac{RAN} slices, each of them customized to meet diverse \ac{QoS} requirements. 
A RAN soft-slicing approach based on network-level resource pre-allocation is proposed in \cite{RANSliceFramework_prescheduling} to enable opportunistic resource sharing among slices and support instantaneous service demands.
With the aim of improving resource utilization efficiency, the RAN slicing control strategy in \cite{RANSliceFramework_ai} adopts multiple time-resource granularities, where radio resources can be dynamically shared between slices.
However, the potential of \ac{RAN} slicing on addressing the inherent smart grid service heterogeneity has not yet been adequately explored in the literature.
Focusing on smart grid service provisioning, this article} introduces a beyond-5G \ac{RAN} slicing framework using the IEC 61850 standard to define smart grid communication requirements. 
In summary, our contribution is threefold:
\begin{enumerate}%
    \item We present a comprehensive categorization of 5G-enabled smart grid services, highlighting the role of \ac{RAN} slicing on \comt{their efficient integration}.%
    \item We propose a \comt{novel \ac{RAN} slicing framework empowered by \ac{AI} for the accommodation of IEC 61850 services in beyond-5G systems}.%
    \item \comt{We demonstrate the feasibility of our approach in a smart grid self-healing use case, where efficient radio resource allocation is achieved while conforming to peculiar \ac{QoS} requirements}.%
\end{enumerate}%

The rest of the paper is organized as follows: 
Section \ref{sec:Two} presents a classification of smart grid services enabled by \ac{5G} systems \comt{and their associated slices.}
Section \ref{sec:Three} describes \comt{key} aspects of RAN slicing and highlights the major benefits for smart grid communication from technological and business perspective.
Section \ref{sec:Four} \comt{introduces} our proposed RAN slicing framework \comt{for IEC 61850 services, with a comprehensive description of its building blocks.}
Section \ref{sec:Five} \comt{outlines our AI-based methodology and presents a performance assessment pertaining to a smart grid communication scenario.}
\comt{Finally, concluding remarks are summarized in Section \ref{sec:Six}.}

\section{Classification of 5G-enabled Smart Grid Services} \label{sec:Two}
Two interdependent domains form the smart grid infrastructure: \textit{i)} a hierarchical power system, covering multi-directional power flow \comt{steps from generation to final consumption;} and \textit{ii)} a two-way communication system, enabling \comt{extensive} information exchange.
\comt{In what follows, we provide a categorization of 5G-enabled smart grid services and we highlight the relationship with 5G network slices based on their requirements.}

\vspace{-.3cm}
\subsection{Smart Distribution Automation}
Distribution automation allows power distribution systems to reconfigure themselves when a fault occurs, restricting the problem to a smaller area~\cite{HO13}. 
Rapid fault location, isolation, and service restoration offered by \acp{IED} and other controllable units reduce the total outage time and the number of interruptions.
\comt{5G-enabled distribution automation} aims to achieve real‐time situational awareness and quasi‐real-time analysis of the grid behaviour by supporting advanced functionalities, e.g., automated feeder switching and optimized restoration dispatch. \comt{Such operations are often linked to stringent performance requirements, i.e., very low end‐to‐end latency and ultra‐high reliability, falling under the \ac{uRLLC} network slice.}
\vspace{-.3cm}
\subsection{Wide-Area Monitoring, Control, and Protection}
\comt{Power system} operators install sensors on critical grid components, such as power lines and transformer banks, to measure equipment status parameters.
\comt{Such} measurements provide real-time alerts for abnormal conditions and outage information to support utilities in predicting equipment maintenance and replacement. 
By exploiting \ac{5G} and beyond communication infrastructure, wide-area monitoring systems aim to enhance traditional supervisory control and data acquisition (SCADA) systems, offering advanced supervision \comt{capabilities} \cite{8067687}. 
\comt{Since the characteristics of the monitoring elements vary, this service category requires a synergistic mix of \ac{uRLLC}, \ac{mMTC}, and \ac{eMBB} slices}.
\vspace{-.3cm}
\subsection{Metering Data Acquisition}
The massive deployment of smart meters for large-scale information acquisition constitutes one of the \comt{principal} components of next-generation power systems \cite{MEN14}.
Smart meters continuously evolve to sophisticated computing units, which gather, process, and transmit user consumption information to data aggregation units for further processing and analysis.
\comt{The advanced \ac{mMTC} capabilities of beyond-5G systems are instrumental for generating unprecedented volumes of metering information.}
In addition, \comt{advanced metering} systems can leverage the distributed information processing and storage architecture of 5G and beyond core networks, supporting fog computing platforms for localized decision-making.
\vspace{-.3cm}
\subsection{Distributed Generation Integration and Microgrids}
By providing higher fault tolerance and islanding detection, 
\comt{5G-enabled smart grids} enable safer and more reliable connections of distributed generation units, e.g., solar photovoltaic panels, wind turbines, and natural-gas-powered fuel cells. 
The increasing penetration of renewable energy sources gives rise to the microgrid paradigm, which acts as a single controllable entity concerning the grid, i.e., operation in either a grid-connected or island mode.
Integrating intermittent renewable sources and microgrid management require advanced control techniques and networking schemes for seamless \comt{operation}, especially when the energy storage capacity is low. 
\comt{The diverse requirements introduced by this service category are a combination of traditional \ac{uRLLC} and \ac{mMTC} network slices.}

\vspace{-.3cm}
\subsection{Volume and Price Balancing}
The introduction of smart grid has pushed the roll-out of demand response programs with flexible management of energy consumption at consumer ends in response to supply conditions regulated by the utility providers~\cite{5G_microperators_ahokangas}.
Through extensive information exchange provided by 5G networks, energy consumers can be transformed into prosumers, who interact and collaborate by producing, consuming, storing, and exchanging energy on a peer-to-peer basis. 
\comt{Such} decentralized energy optimization strategies, performed locally, allow markets to determine prices accurately, resulting in cost savings. 
\comt{Considering a high number of prosumers and market information exchanges with historical load data, this service category requires a combination of \ac{mMTC} and \ac{eMBB}}.

Table~\ref{table:requirements_servicescenarios} summarizes the categorization of 5G-enabled smart grid services with their associated network slice(s), indicated by the standardized \ac{SST} index. It is worth noting that each \ac{SST} may be linked to diverse requirements within the same slice\comt{, with the aid of a slice differentiator parameter which builds customized fine-grained slices.} 
%

\vspace{-.25cm}
\section{RAN Slicing for the Smart Grid}\label{sec:Three}

This section \comt{highlights key} aspects of \ac{RAN} slicing, underlying the benefits of \comt{integrating the multi-faceted connectivity landscape of smart grid services,}
from both technological and business perspectives.

\vspace{-.38cm}
\subsection{Key Features} \label{sec:ThreeA}
\begin{figure}[t!]
   \centering
   \includegraphics[width=0.5\textwidth]{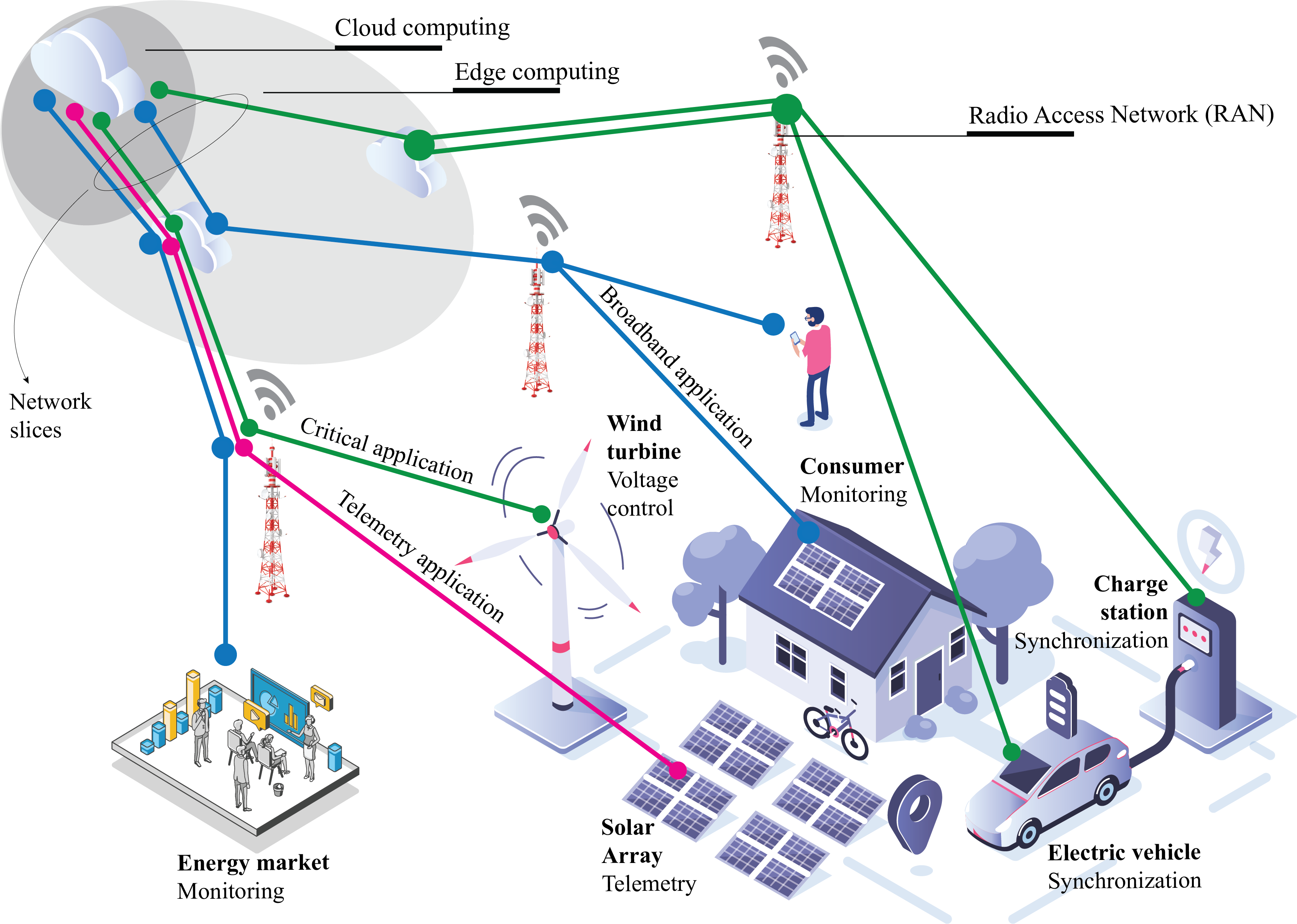}
   \caption{RAN slicing supports a variety of smart grid \comt{services}, creating multiple isolated logical networks on top of an underlying physical infrastructure.}
   \label{fig:networslices}
   \vspace{-.4cm}
 \end{figure}

In stark contrast to previous mobile network generations, the 5G architecture leverages cloud-native concepts to allow the disaggregation and virtualization of network functions \cite{3GPP.23.501}.
\comt{By leveraging virtualized network function management, 
smart grid services can be instantiated in a flexible manner, following a network-as-a-service model.}
%
%
\comt{This architectural shift towards network softwarization allows differentiated handling of smart grid traffic types, and ensures their harmonic coexistence with proper allocation of storage, computing, processing and radio resources.}
A slice-based \ac{5G} RAN can efficiently address the inherent smart grid service heterogeneity, as illustrated in Fig.~\ref{fig:networslices}\comt{, and} offers significant benefits \comt{summarized as follows:}
\subsubsection{Service isolation}
\comt{Network resources of a smart grid service can be isolated from other service resources in a resilient manner without violating the agreed \acp{SLA}.
In addition to increased security/privacy, service isolation offers a higher reliability potential owing to the guaranteed resources required to achieve stringent performance requirements, e.g., ultra-high reliability for protection-related functions in substation communication.}
\subsubsection{Tenant-oriented virtual network}
\comt{\ac{RAN} slice tenants
are able to operate their own dedicated network and manage the provisioning of customized network components based on the adopted business models.
%
The tenant can be an individual energy consumer or a power system utility that leases and occupies RAN slice instances.
By allowing partial/full control of the virtualized end-to-end networks to tenants, RAN slicing clearly differentiates from other network sharing techniques, offering new roles to energy players.}
\subsubsection{Guaranteed SLA}
\comt{RAN slicing offers differentiated service provisioning with \ac{QoS} guarantees based on negotiated \acp{SLA}. A management and orchestration entity is responsible for mapping the requirements established in \acp{SLA} into the functional elements of RAN slices. Due to the mission-critical nature of certain smart grid services, SLA monitoring needs to be predictive while ensuring efficient radio resource utilization. In this context, the exploitation of emerging \ac{AI} techniques holds the promise of conforming to \acp{SLA} by exploiting slice state information.}
\subsubsection{Customization capabilities}
\comt{The implementation of customized functions is a key RAN slicing feature, given the highly diverse service ecosystem in smart grids. 
By leveraging edge computing, 
decision-making and intelligence can be shifted closer to the network endpoints, achieving 
faster response times and targeted operational actions. Agile function placement facilitates the support of new use cases and often results in reduced installation/operational cost for utilities.}
\subsubsection{Distributed architecture}
\comt{RAN slicing promotes a decentralized structure of the smart grid against the traditional model, in which non-cooperative systems are deployed and managed independently in a hierarchical manner. On-demand deployment of virtual functions facilitates the seamless integration of DERs and paves the way towards a \textit{prosumer-centric} vision of future power systems. Smart grid protection services requiring low latency can be performed autonomously rather than being delegated to a central management unit \cite{8067687}.}

\vspace{-.33cm}
\subsection{Business Potential}
%
The support of 5G-enabled smart grid services via RAN slicing comes along with a continuously rising number of subscriptions and traffic demand, giving rise to significant business opportunities. 
The stakeholders, i.e., the beneficiaries in the network slicing ecosystem, consist of the \ac{NSSI} provider, the intermediate-\ac{NSI} provider, the \ac{E2E-NSI} provider, the slice tenant, and the end customer. 
\comt{In some cases, multiple network slice providers may coexist to provide an E2E slice, while in other business scenarios a single \ac{MNO} may undertake the role of E2E slice provider.}
The service-based business model promoted by \ac{RAN} slicing, motivates \acp{MNO} to move beyond traditional subscription-based schemes with fixed rental fees to more flexible pricing policies \comt{and direct value propositions to the smart grid utilities, e.g., revenue split schemes and incentive strategies.} As a result, the new electricity business models on ownership, operation, maintenance as well as usage, will be defined according to performance-based \acp{SLA} and relevant key performance indicators \comt{of the electrical infrastructure}, which could be complemented by the added value on each utility.
%
%

%
\comt{From the perspective of power system utilities, service provisioning and cost efficiency empowered by RAN slicing, in conjunction with the deregulation of the energy sector, reinforce new business practices.}
The opportunities of this evolving market context are expected to alter the way transmission/distribution system operators use connectivity technologies in the grid. 
\comt{Leveraging \ac{RAN} slicing, novel use cases and diversified requirements can be supported using a unique communication infrastructure with significant cost savings.}

\begin{figure}[t!]
\centering
\includegraphics[width=0.45\textwidth]{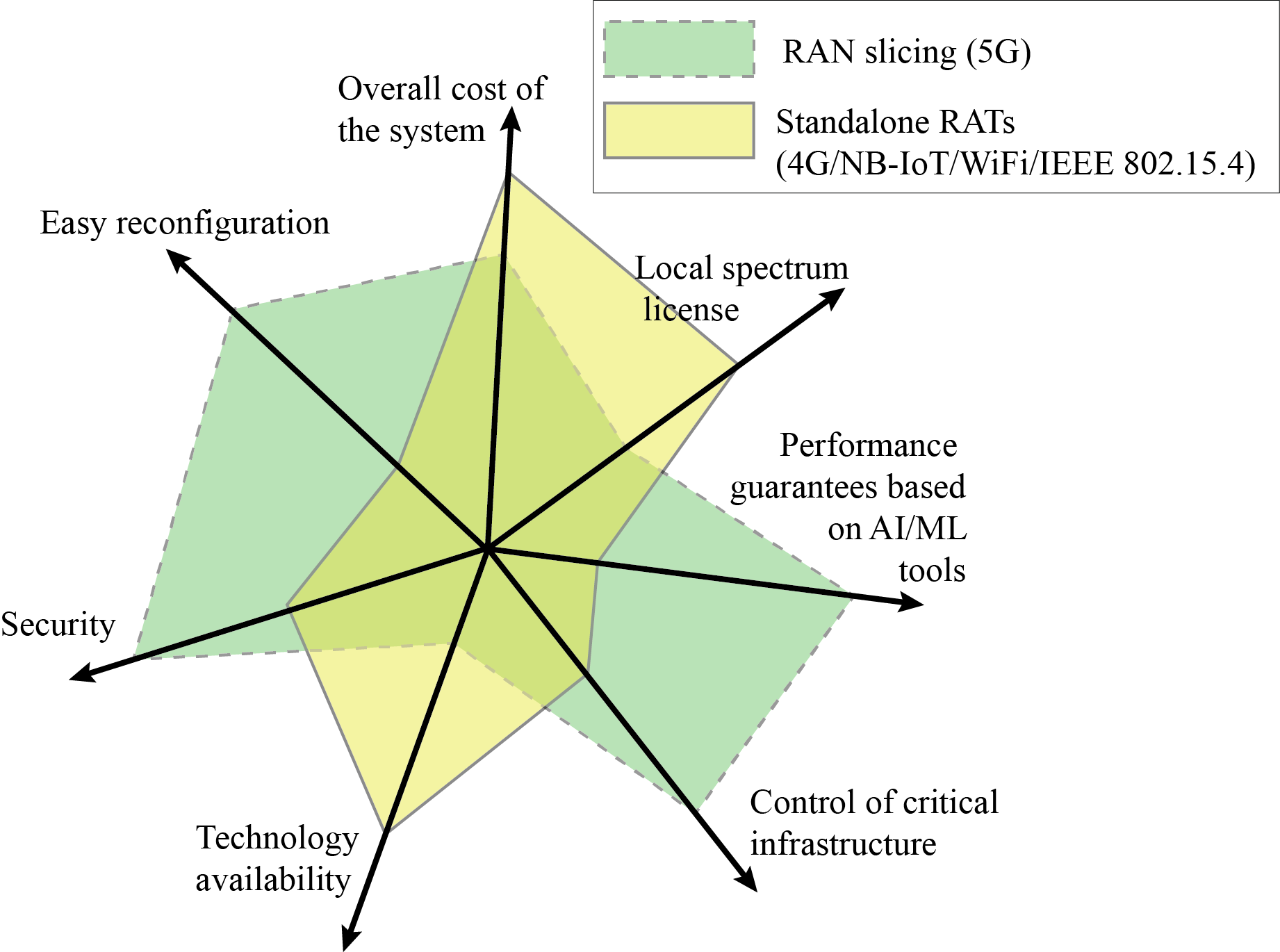}
\caption{Comparison between a RAN slicing framework and \comt{wireless} \acp{RAT}, such as \ac{LTE}, \ac{NB-IoT}, Wi-Fi, and IEEE 802.15.4, \comt{when applied as standalone connectivity solutions for smart grid services}.}
\label{fig_spider}
\vspace{-.35cm}
\end{figure}

A qualitative comparison between \ac{RAN} slicing and existing radio access technologies (RATs) for the smart grid is illustrated in Fig. \ref{fig_spider}. \comt{It is worth noting that} in terms of technology availability, existing RATs have an essential advantage compared with RAN slicing\comt{, related to the maturity of communication standards and the already well-established ecosystems adopting such technologies.} The same applies when local spectrum licenses are \comt{in force.} \comt{Nonetheless, enhanced security mechanisms, reconfiguration capabilities, overall system cost, control of the critical infrastructure, and incorporation of artificial intelligence are some of the aspects where RAN slicing demonstrates clear superiority.} 

\begin{figure*}[ht!]
\centering
\includegraphics[width=0.8\textwidth]{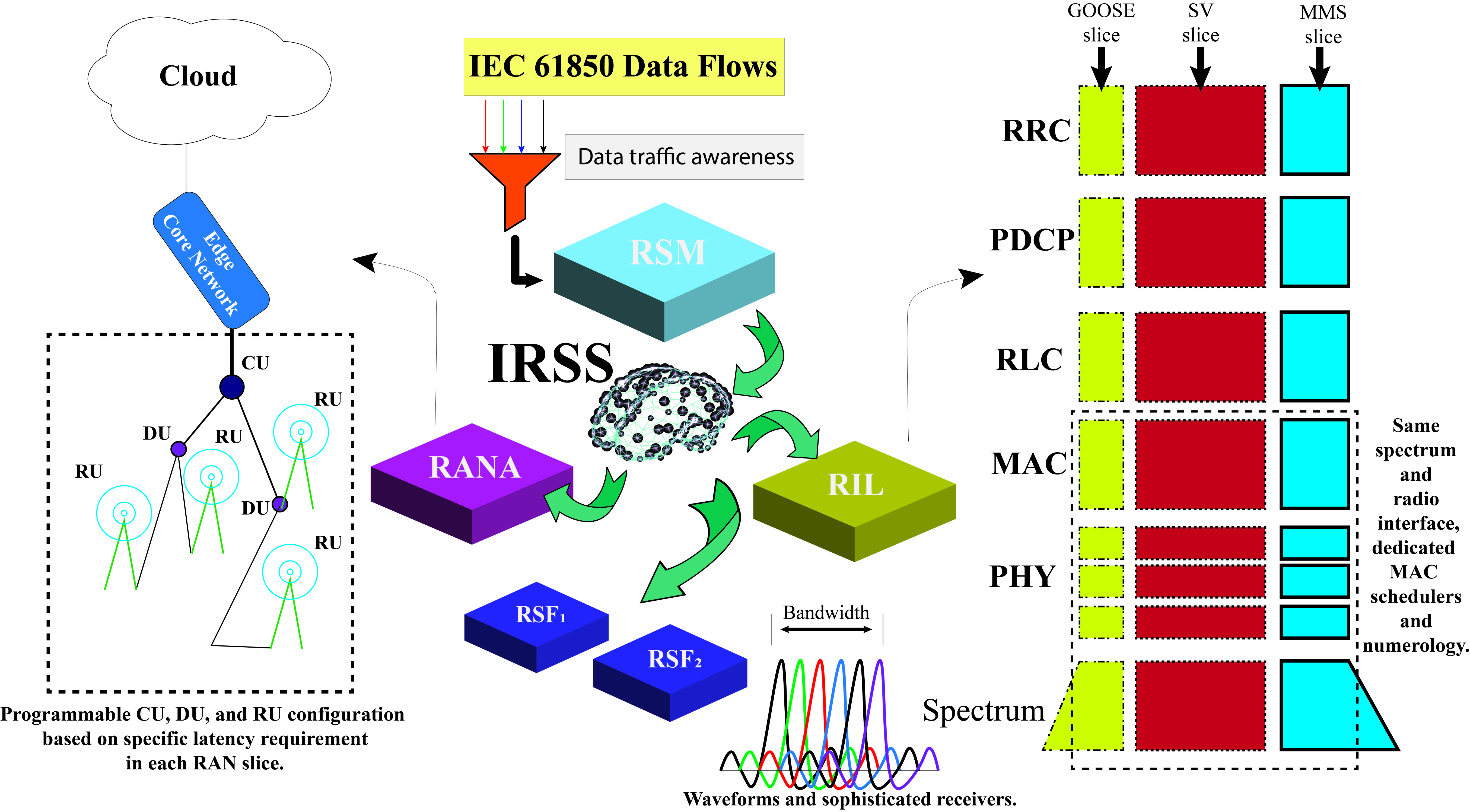}
\caption{A functional RAN slicing framework to support wireless connectivity for IEC 61850 \comt{services}.}
\vspace{-.35cm}
\label{fig_integration_framework}
\end{figure*}

\vspace{-.11cm}
\section{The Proposed Framework}\label{sec:Four}
\comt{The diverse QoS requirements in the smart grid provide a fertile ground for the application of an agile RAN slicing approach.} \comt{Our methodology is devised to cope with the rising complexity of supporting 5G-enabled smart grid services, achieving not only more manageable RAN slices but also conforming to the business propositions sought by network operators and smart grid utilities.} The proposed framework complements the RAN slicing enhancements introduced in \cite{3gpp.38.832}, by bridging the gap between IEC 61850 services and the 3GPP-standardized radio resource sharing strategies.

\vspace{-.11cm}
\subsection{The IEC 61850 Standard}\label{sec:Four-A}
In terms of power industry communication standards, the IEC 61850 standard is of particular note. \comt{Originally defined} to cover the stringent requirements for automation within \comt{electrical} substations, IEC 61850 emerges as a versatile interoperable standard that can be applied beyond the substation boundaries to facilitate intersubstation message exchange, wide-area transmission of synchrophasor information \comt{and} DERs' communication. IEC 61850 promotes abstract and application-specific data models and services that are decoupled from the underlying communication technologies. As the applicability of the standard expands continuously, there is growing research interest to integrate IEC 61850 services with wireless protocols and overcome the physical limitations and high installation costs of the default Ethernet technology.

The IEC 61850 standard defines the performance classes and communication requirements for various message types exchanged between power system components. In this context, the IEC 61850 message classification can be intrinsically associated with the RAN slicing concept. In particular, \comt{Type-1 and Type-6 messages impose requirements that fall under the \ac{uRLLC} slice, as they are linked to real-time substation protection actions and time synchronization, respectively.} For instance, in line phase comparison for analog protection, Type-1 messages must be delivered with ultra-high reliability levels, i.e., a packet loss rate in the order of $10^{-5}$ \cite{IEC_requirements}. Other message types, related to continuous IED data streams (or large file transfers) and sporadic medium-speed event reports, can be mapped to \ac{eMBB} and \ac{mMTC} slices, respectively.

\vspace{-.25cm}
\subsection{RAN Slicing for IEC 61850 Services}\label{sec:Four-B}
Our RAN slicing framework aims to accommodate IEC-61850 services over a single shared infrastructure and lays the foundation for a fine-grained service management in the smart grid. As illustrated in Fig. \ref{fig_integration_framework}, \comt{it is primarily composed of a number of interworking functional components with programmable capabilities, aiming at a flexible instantiation of IEC 61850 services.}
\comt{In what follows, we provide a concise description of the building elements defined for RAN slicing.}

\vspace{0.03in}
\subsubsection{\ac{RANA}}
\comt{The next-generation NodeB (gNodeB) is a key component in the RAN slicing architecture.
It provides \ac{RAN} slice subnets which consist of the \ac{CU}, multiple \acp{DU} and multiple \acp{RU}.}
Heterogeneous IEC 61850 requirements can be supported by appropriate \ac{RAN} operating principles that involve different \comt{functional} roles among the \acp{RU}, the \acp{DU} and the \ac{CU}. Such roles are determined based on the \ac{RANA} index value\comt{, offering \ac{QoS} flexibility to \acp{MNO} to select the appropriate deployment option.}
%
\comt{For certain \ac{RANA} values},
the transmission of IEC 61850 message types follows 
a centralized (i.e., client/server) architecture to support \comt{traffic types} with \comt{moderate} end-to-end latency levels.
On the other hand, \comt{proper \ac{RANA} configurations} may also enable
a (fully) distributed architecture that relies on \comt{a disaggregated functional split to} \acp{DU} and/or \acp{RU}. In this case, \comt{processing functions are primarily located closer to the \acp{DU} and/or \acp{RU},} achieving lower latency levels for
critical IEC 61850 messages, e.g., IEC 61850 \comt{traffic types} with $\leq$10ms or $\leq$3ms latency budgets \cite{IEC_requirements}.

\vspace{0.03in}
\subsubsection{\ac{RIL}}
Considering that isolation \comt{represents the state in which the performance degradation of one slice does not impact the performance of other slices}
\cite{RAN_slicing_verticals_eddine}, the \ac{RIL} functional element defines the required isolation level among slices, which reflects a trade-off between isolation and system efficiency.
\comt{In an inter-slice sharing strategy, radio resource management for IEC 61850 messages reduces the associated slice tenant costs. On the other hand, a conservative isolation policy employs
dedicated radio resource assignments, conforming to stringent security and privacy enforcements.}

\vspace{0.03in}
\subsubsection{\ac{RSF}}
This building element is mainly associated with the functionalities performed in the \comt{\acp{DU}}.
\comt{To ensure a future-proof element design, a partition of these functionalities is carried out into two well-defined subelements.}
The RSF$_1$ comprises the already standardized \ac{5G}/NR functionalities and involves the scalable \ac{5G}/NR numerology on the frame structure \cite{RAN_slicing_verticals_eddine}.
This subelement is associated with fundamental \ac{RAN} functions, such as tiling, scheduling and puncturing.
Tiling refers to the assignment of radio resources into different tiles defined in the time/frequency space, which can be individually configured according to the IEC 61850 service requirements and the respective \acp{SLA}.
Scheduling refers to the allocation of radio resources, while traffic puncturing allows efficient multiplexing of various IEC 61850 services by prioritizing the transmission of time-sensitive message types.
On the other hand, the RSF$_2$ subelement consists of customized baseband functionalities to support vendor-specific operations or features \comt{not yet specified} by standards.
It considers beyond-\ac{5G} technical enhancements, paving the way for next-generation connectivity enablers to be incorporated in the RAN slicing framework. 

\vspace{0.03in}
\subsubsection{\ac{RSM}}
This \comt{component} monitors the necessary capabilities of the aforementioned \ac{RANA}, \ac{RIL} and \ac{RSF} elements.
With a supervisory and management role in the RAN domain, the \ac{RSM} \comt{handles RAN slice instantiation and lifecycle management of IEC 61850 services.}
The \ac{RSM} \comt{interacts} with the \textit{data traffic awareness} \comt{module}, a specialized function used to \comt{process}, analyze and evaluate the IEC 61850 \comt{data flows}.
\comt{Its intent-driven operation} needs to rigorously consider the diversified \acp{SLA} of IEC 61850 services designated by the core network.
\comt{The fundamental hallmark of RSM element is therefore to guarantee a harmonic co-existence of multiple RAN slices and their respective \acp{SLA}.}

\vspace{0.06in}
\subsubsection{\ac{IRSS}} 
This element adds cognition to the proposed framework.
\comt{The key operation of \ac{IRSS} lies in the knowledge extraction from the aggregated IEC 61850 traffic for \ac{RAN} slice scheduling and radio resource assignment. The exploitation of advanced \ac{AI} techniques by the \ac{IRSS} holds the promise of achieving a high degree of automation and operational efficiency for RAN slicing. Radio resource allocation in \ac{IRSS} needs to ensure an efficient multiplexing of IEC 61850 services by prioritizing the transmission of time-sensitive message types. To ensure such contextual decision-making, the \ac{IRSS} interacts with all aforementioned elements of the framework for \ac{RAN} slice awareness, e.g., \ac{SLA} monitoring and resource isolation.}
%

\vspace{-.3cm}
\section{Performance Assessment}\label{sec:Five}
\subsection{Network Scenario}

\begin{figure}[!t]
\centering
\includegraphics[width=0.5\textwidth]{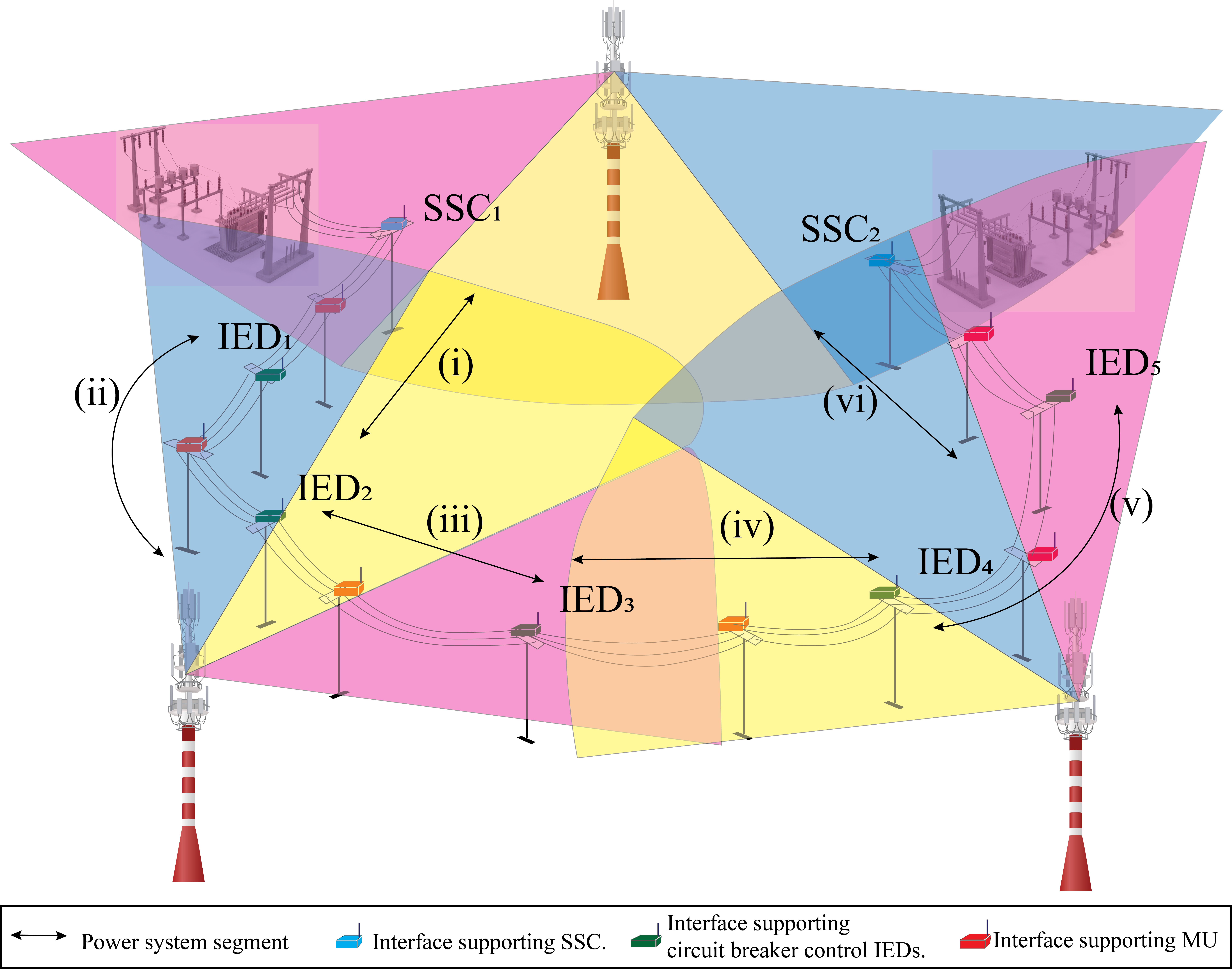}
\caption{\comt{Smart grid self-healing use case with a cellular network deployment} 
supporting IEC 61850 \comt{services}. \comt{The gNodeBs} \comt{provide connectivity to different segments of the power system denoted as (i)-(vi)}. The \acfp{SSC}, \acfp{IED}, and \acfp{MU} \comt{are equipped with 5G interfaces, and exchange system-related information by sharing available radio resource blocks}.}
\vspace{-0.5cm}
\label{fig_integration_framework_2}
\end{figure}

To evaluate the \comt{feasibility} of the proposed \ac{RAN} slicing framework, we consider the smart grid self-healing scenario in~\cite{self_healing_scenario}, where
automatic reconfiguration occurs after a short-circuit \comt{fault}.
The \acp{IED}, \acp{SSC}, and \acfp{MU} are equipped with \ac{5G} interfaces to transmit \comt{IEC 61850 messages} in a multicell network topology.
As illustrated in Fig. \ref{fig_integration_framework_2}, \comt{connectivity is provided by gNodeBs deployed to provide coverage in the different segments of the power system delimited by the \acp{IED} and \acp{SSC}.}
Three \ac{RAN} slices are considered, corresponding to the \ac{GOOSE}, \ac{SV}, and \ac{MMS} services in IEC 61850. In particular:
\begin{enumerate}
    \item \textbf{\ac{GOOSE}} slice, \comt{supporting \ac{uRLLC}-type information exchange between \acp{IED}, to isolate the fault and restore power supply in other power system segments. Short transmission latency of \ac{GOOSE} messages is critical to minimize the impact on power system stability.}
    \item \textbf{\ac{SV}} slice, supporting \comt{\ac{uRLLC}-type} information exchange between \acp{MU} and \acp{IED}. \comt{SV messages contain current and voltage measurements with accurate timestamp data. Similar to \ac{GOOSE}, stringent latency requirements apply.}
    \item \textbf{\ac{MMS}} slice, supporting \comt{\ac{mMTC}-type} information exchange between \acp{IED} and \comt{SCADA centers, to monitor the power system segments}. \comt{Such messages have relatively moderate latency requirements compared to \ac{GOOSE} and \ac{SV} slices.}
\end{enumerate}
Path loss model, radio configurations and simulation assumptions follow the 3GPP specifications in \cite{3gpp.36.931}. \comt{Radio resource blocks are shared among the active IEC 61850 communication links; thus, each smart grid device suffers from  co-channel interference caused by other devices sharing the same resource blocks.}
Dynamic scheduling \comt{for RAN slices} is necessary to efficiently manage resource allocation, minimize interference, and prioritize specific IEC 61850 \comt{messages according to \acp{SLA}.}

\vspace{-.2cm}
\subsection{IRSS Implementation}
\comt{To address the joint radio resource assignment and power control problem for smart grid devices, the \ac{IRSS} employs two deep reinforcement learning (DRL) algorithms which collaboratively aim to maximize the achieved spectral efficiency of all active IEC 61850 communication links. DRL models sequential decision-making problems with an agent and an environment interacting and exchanging information in the form of \textit{states}, \textit{actions}, and \textit{rewards} \cite{sutton2018reinforcement}. 
With the aid of deep neural networks for function approximation, smart grid devices successively learn two \textit{policies} to determine their assigned radio resources and transmit power levels, and 
maximize the expected sum of rewards. The reward function, common for both algorithms, takes into account \textit{i}) the achieved spectral efficiency of each smart grid device and \textit{ii}) the interference level caused to other devices.}
\begin{figure}
\begin{tabular}{c}
       \psfrag{p}{\tiny$\Omega$}
\psfrag{g}{\tiny$\theta$}
\psfrag{h}{\tiny$\gamma$}
\psfrag{k}{\tiny$\Omega_\text{agent}$}
\psfrag{b}{\tiny$\theta_\text{agent}$}
\psfrag{e}{\tiny update $\Omega$}
\psfrag{t}{\tiny update $\theta$}
 \includegraphics[width=0.45\textwidth]{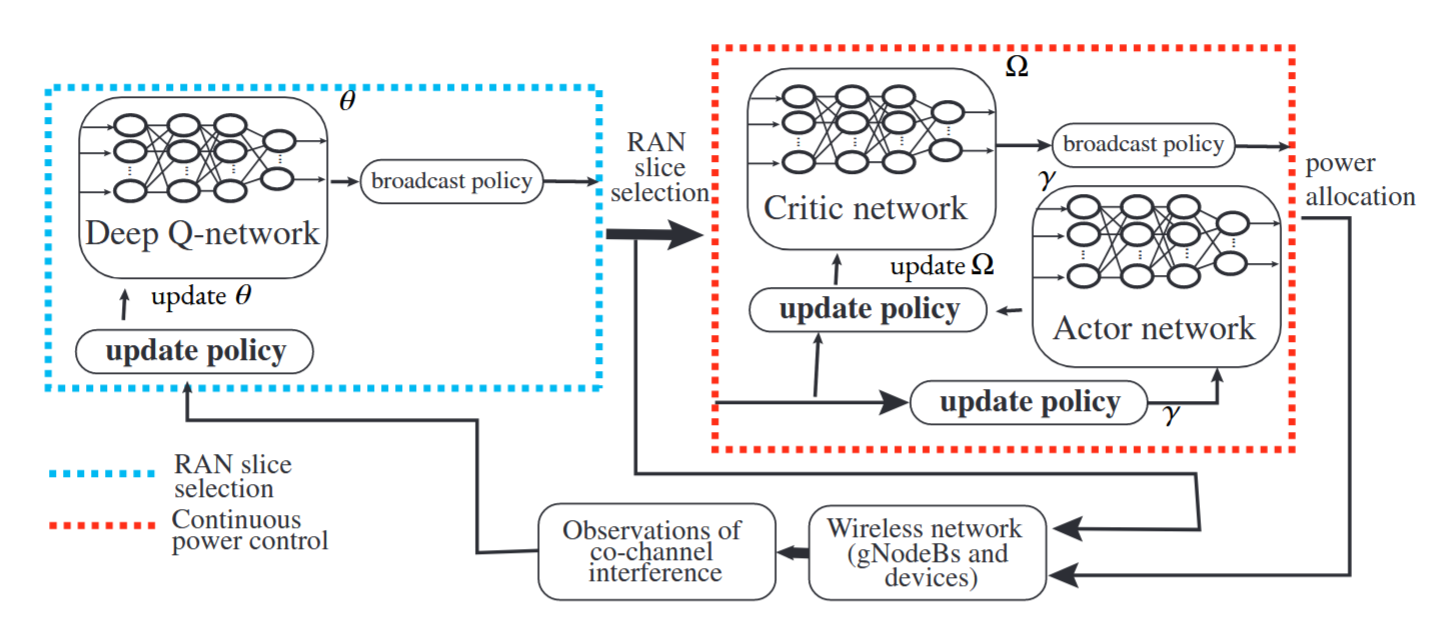}\\
       (a) \\
 \includegraphics[width=0.45\textwidth]{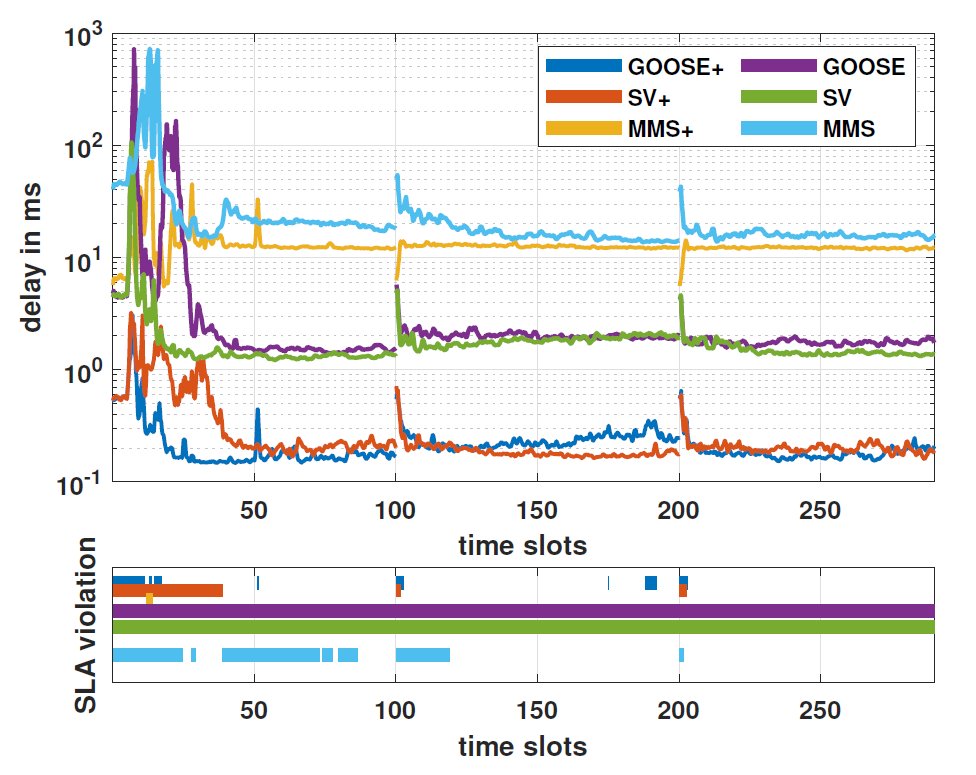}\\
     (b)
\end{tabular}
 \caption{(a) \comt{\ac{DRL}-based \ac{IRSS} implementation for radio resource assignment and power control of smart grid devices. The DQN layer provides the resource assignment decision to the actor-critic algorithm, which, in turn, determines the transmit power levels. (b) \ac{RAN} slice performance assessment for three IEC 61850 services in terms of latency and \ac{SLA} violation rate in a smart grid self-healing use case. For slices labeled with symbol +, \ac{DRL}-based \ac{IRSS} takes into account the beyond-5G configuration options for RANA, RIL and RSF elements.} Latency requirement for time-sensitive \ac{GOOSE} and \ac{SV} services is set to 0.3 ms and 0.5 ms, respectively, whereas a latency budget of 40 ms is specified for \ac{MMS} services \cite{IEC_requirements}.}
 \vspace{-0.8cm}
   \label{figura_BER1}
\end{figure}
\comt{As illustrated in Fig.~\ref{figura_BER1}a, IRSS employs the following two learning layers:}

\begin{itemize}
    \item \comt{A deep \textit{Q}-network (DQN) algorithm is considered for resource assignment to the different RAN slices. DQN is a value-based algorithm, applicable to environments with discrete action space. To strike a balance between exploring state-action pairs and exploiting knowledge, we adopt a Boltzmann policy to steer exploration towards more promising actions off the \textit{Q}-value-maximizing path, instead of selecting all actions with equal probability \cite{sutton2018reinforcement}. During training, experiences gathered by older policies are stored in a replay memory, and are reused to improve sample efficiency.}
    \item \comt{An actor-critic algorithm is applied to manage the continuous action space for transmit power allocation. In actor-critic schemes, two components are learned jointly; an actor, which learns a parameterized policy, and a critic which learns a value function to evaluate state-action pairs. The actor first receives as input the resource block assignment decision from the DQN layer. Using a learned value function, the critic provides a reinforcing signal to the actor which can be more informative for a policy than the rewards from the environment. In our method, we learn an \textit{advantage} function as the reinforcing signal, which measures the extent to which an action is better or worse than the policy's average action in a state. The estimation of advantage function is performed using an exponentially weighted average of a number of advantage estimators with different bias and variance \cite{schulman}.}
\end{itemize}

\vspace{-0.4cm}
\subsection{Results}
\comt{The top plot in Fig. \ref{figura_BER1}b illustrates the latency performance for \ac{GOOSE}, \ac{SV} and \ac{MMS} slices for two different simulation setups. In particular, the transmission latency
for each IEC 61850 communication link is measured as the ratio of the packet size (prescribed in \cite{IEC_requirements}) over the
achieved throughput. In the first setup (i.e., slices labeled with symbol +), the DRL-based \ac{IRSS} element takes into consideration the beyond-5G configuration options for \ac{RANA}, \ac{RIL} and \ac{RSF} elements, as described in Section \ref{sec:Four}. In this case, the augmented state-action space for each smart grid device opens up the possibility of discovering better states and ways of acting in the quest for the optimal resource and power decisions. In the second setup, the DRL-based \ac{IRSS} element uses default 5G configurations for resource assignment and power control according to \cite{3gpp.36.931}. It can be observed that \ac{IRSS} achieves superior latency performance in the first setup compared to the second one, while conforming to the \ac{SLA} requirements for \ac{GOOSE}, \ac{SV} and \ac{MMS} slices as specified by the \ac{RSM} element. 
The considered \acp{SLA} refer to the latency requirements for IEC 61850 services, and 
an \ac{SLA} violation occurs when the achieved
latency level becomes higher than the threshold value~\cite{IEC_requirements}.
In addition, \ac{IRSS} learns to prioritize time-sensitive \ac{GOOSE} and \ac{SV} services, resulting in significant latency reduction compared to the default 5G configuration. This, in turn, leads to fewer \ac{SLA} violations, as illustrated in the bottom plot of Fig. \ref{figura_BER1}b.}

\comt{We note that the transient latency behavior for all \ac{RAN} slices identified in the first time slots is attributed to the exploration phase of the DRL-based \ac{IRSS}, which may result in suboptimal actions by each smart grid device. However, as training progresses, the rate of exploration gradually decays, and smart grid devices learn better policies for their assigned resources and transmit power levels.}

\vspace{-0.2cm}
\section{Concluding Remarks} \label{sec:Six}
The smart grid paradigm undoubtedly represents an essential showcase for 5G and beyond systems, mainly because of the heterogeneous connectivity landscape and the wide range of service requirements. \comt{In this context, RAN slicing poses elevated merit for a fine-grained smart grid service management with a shared communication infrastructure. At the same time, the proliferation of AI techniques for sequential decision-making problems offers remarkable benefits in RAN slice management.}
This article \comt{introduced an AI-native RAN slicing framework for the integration of IEC 61850 services in beyond-5G systems. Our proposed framework comprises multiple functional elements which interoperate towards intent-driven RAN slice management.
The feasibility of our approach was demonstrated with the aid of a smart grid self-healing scenario with diversified QoS requirements. By resorting to two DRL-based algorithms, the joint resource allocation, interference minimization, and IEC 61850 message prioritization can be efficiently handled, while conforming to SLAs.}

\comt{In the path forward, we will direct our efforts towards the design of traffic forecasting modules for IEC 61850 services, and their incorporation in our RAN slicing framework. This will facilitate predictive slice provisioning by ensuring traffic-aware admission control policies. The introduced signaling overhead for \ac{RAN} slicing control will also be quantified.}

\section*{Acknowledgment}
This paper is partly supported by Academy of Finland via: (a) FIREMAN consortium CHIST-ERA/n.326270; (b) EnergyNet Research Fellowship n.321265/n.32886/n.352654; (c) X-SDEN project n.349965; (d) 6G Flagship (n.346208). This research is also supported by the joint Baltic-Nordic Energy Research programme project “Guidelines for Next Generation Buildings as Future Scalable Virtual Management of MicroGrids [Next-uGrid]” (n.117766), and Finnish public funding agency for research, Business Finland under projects 5GVIIMA and IFORGE, the projects are parts of 5G Test Network Finland (5GTNF) and Smart Otaniemi ecosystems. The work of Charalampos Kalalas was supported by FIREMAN project CHIST-ERA-17-BDSI-003 funded by the Spanish National Foundation (PCI2019-103780). The work of Petar Popovski was, in part, supported by the Villum Investigator Grant “WATER” from the Velux Foundation, Denmark.

\ifCLASSOPTIONcaptionsoff
  \newpage
\fi

\vspace{-.3cm}
\bibliographystyle{IEEEtran}
\bibliography{references}

\begin{thebibliography}{10}
\providecommand{\url}[1]{#1}
\csname url@samestyle\endcsname
\providecommand{\newblock}{\relax}
\providecommand{\bibinfo}[2]{#2}
\providecommand{\BIBentrySTDinterwordspacing}{\spaceskip=0pt\relax}
\providecommand{\BIBentryALTinterwordstretchfactor}{4}
\providecommand{\BIBentryALTinterwordspacing}{\spaceskip=\fontdimen2\font plus
\BIBentryALTinterwordstretchfactor\fontdimen3\font minus
  \fontdimen4\font\relax}
\providecommand{\BIBforeignlanguage}[2]{{%
\expandafter\ifx\csname l@#1\endcsname\relax
\typeout{** WARNING: IEEEtran.bst: No hyphenation pattern has been}%
\typeout{** loaded for the language `#1'. Using the pattern for}%
\typeout{** the default language instead.}%
\else
\language=\csname l@#1\endcsname
\fi
#2}}
\providecommand{\BIBdecl}{\relax}
\BIBdecl

\bibitem{GUN11}
V.~{Gungor} \emph{et~al.}, ``{Smart Grid Technologies: Communication
  Technologies and Standards},'' \emph{IEEE Trans. Ind. Informat.,}, vol.~7,
  no.~4, pp. 529--539, Nov. 2011.

\bibitem{HO13}
Q.-D. {Ho} \emph{et~al.}, ``{Challenges and research opportunities in wireless
  communication networks for smart grid},'' \emph{IEEE Wireless Commun.,},
  vol.~20, no.~3, pp. 89--95, Jun. 2013.

\bibitem{8067687}
M.~{Cosovic} \emph{et~al.}, ``{5G Mobile Cellular Networks: Enabling
  Distributed State Estimation for Smart Grids},'' \emph{IEEE Commun. Mag.},
  vol.~55, no.~10, pp. 62--69, 2017.

\bibitem{5G_microperators_ahokangas}
P.~{Ahokangas} \emph{et~al.}, ``{Business Models for Local 5G Micro
  Operators},'' \emph{IEEE Trans. Cogn. Commun. Netw.}, vol.~5, no.~3, pp.
  730--740, 2019.

\bibitem{RANSliceFramework_prescheduling}
J.~Li, W.~Shi, P.~Yang, Q.~Ye, X.~S. Shen, X.~Li, and J.~Rao, ``A hierarchical
  soft ran slicing framework for differentiated service provisioning,''
  \emph{IEEE Wireless Commun.}, vol.~27, no.~6, pp. 90--97, 2020.

\bibitem{RANSliceFramework_ai}
J.~Mei, X.~Wang, K.~Zheng, G.~Boudreau, A.~B. Sediq, and H.~Abou-Zeid,
  ``Intelligent radio access network slicing for service provisioning in 6g: A
  hierarchical deep reinforcement learning approach,'' \emph{IEEE Transactions
  on Communications}, vol.~69, no.~9, pp. 6063--6078, 2021.

\bibitem{MEN14}
W.~{Meng} \emph{et~al.}, ``{Smart grid neighborhood area networks: a survey},''
  \emph{IEEE Netw.}, vol.~28, no.~1, pp. 24--32, Jan. 2014.

\bibitem{3GPP.23.501}
3GPP, ``{Technical Specification Group Services and System Aspects; System
  architecture for the 5G System (5GS); Stage 2},'' Technical Specification
  (TS) 23.501, Mar. 2020, version 16.4.0.

\bibitem{3gpp.38.832}
------, ``{Technical Specification Group RAN; NR; Study on enhancement of Radio
  Access Network (RAN) slicing},'' {3rd Generation Partnership Project (3GPP)},
  Technical Specification (TS) 38.832, Jun. 2021, version 17.0.0.

\bibitem{IEC_requirements}
``\textit{Communication Networks and Systems in Substations -- Part 5:
  Communication Requirements for Functions and Device Models},'' IEC-61850,
  Standard, 2004.

\bibitem{RAN_slicing_verticals_eddine}
S.~E. {Elayoubi} \emph{et~al.}, ``{5G RAN Slicing for Verticals: Enablers and
  Challenges},'' \emph{IEEE Commun. Mag.}, vol.~57, no.~1, pp. 28--34, 2019.

\bibitem{self_healing_scenario}
R.~Ricart-Sanchez, A.~C. Aleixo, Q.~Wang, and J.~M. Alcaraz~Calero,
  ``{Hardware-Based Network Slicing for Supporting Smart Grids Self-Healing
  over 5G Networks},'' in \emph{2020 IEEE International Conference on
  Communications Workshops (ICC Workshops)}, 2020, pp. 1--6.

\bibitem{3gpp.36.931}
3GPP, ``{Technical Specification Group Radio Access Network; Evolved Universal
  Terrestrial Radio Access (E-UTRA); Radio Frequency (RF) requirements for LTE
  Pico Node B},'' {3rd Generation Partnership Project (3GPP)}, Technical Report
  36.931, 03 2022, version 17.0.0.

\bibitem{sutton2018reinforcement}
R.~S. Sutton and A.~G. Barto, \emph{Reinforcement learning: An
  introduction}.\hskip 1em plus 0.5em minus 0.4em\relax MIT press, 2018.

\bibitem{schulman}
\BIBentryALTinterwordspacing
J.~Schulman, P.~Moritz, S.~Levine, M.~Jordan, and P.~Abbeel, ``High-dimensional
  continuous control using generalized advantage estimation,'' 2015. [Online].
  Available: \url{https://arxiv.org/abs/1506.02438}
\BIBentrySTDinterwordspacing

\end{thebibliography}
\vspace{-1cm}
\begin{IEEEbiographynophoto}{Dick Carrillo}[S'01, M'06]
(dick.carrillo.melgarejo@lut.fi) received the B.Eng. degree (Hons.) in electronics and electrical engineering from San Marcos National University, Lima, Per\'u, and the M.Sc. degree in electrical engineering from Pontifical Catholic University of Rio de Janeiro, Rio de Janeiro, Brazil, in 2004 and 2008, respectively. Between 2008 and 2010, he contributed to WIMAX (IEEE 802.16m) standardization. From 2010 to 2018, he worked with the design and implementation of cognitive radio networks and projects based on 3GPP technologies. Since 2018 he is a researcher at Lappeenranta--Lahti University of Technology, where he is also pursuing the doctoral degree in electrical engineering. His research interests are mobile technologies beyond 5G, energy harvesting, intelligent meta-surfaces, cell-free mMIMO, and RAN Slicing. Since 2022, he is a Senior Standardization Specialist at Nokia Bell Labs, where he is contributing on shaping the 3GPP release 18 standard (5G-Advanced).
\end{IEEEbiographynophoto}
\begin{IEEEbiographynophoto}{Charalampos Kalalas}[S'15, M'18]
(ckalalas@cttc.es) received the Ph.D. degree (Cum Laude) in Signal Theory and Communications from the Technical University of Catalonia (UPC) in 2018. He holds an Electrical and Computer Engineering degree (2011) from the National Technical University of Athens (NTUA), Greece, and a M.Sc. degree in Wireless Systems (2014) from the Royal Institute of Technology (KTH), Sweden. He is currently a Researcher with the Sustainable Artificial Intelligence research unit at the Centre Tecnològic de Telecomunicacions de Catalunya (CTTC/CERCA).
\end{IEEEbiographynophoto}
\begin{IEEEbiographynophoto}{Petra Raussi}
(petra.raussi@vtt.fi) received the M.Sc. degree in electrical engineering from LUT University, Lappeenranta, Finland in 2018. She is currently a doctoral candidate at the Department of Electrical Engineering and Automation, School of Electrical Engineering, Aalto University, Espoo, Finland. In 2016, she was a Research Assistant with the School of Energy Systems, LUT University, Lappeenranta, Finland. Since 2017, she has been first a Research Trainee, a Research Scientist, and currently a Senior Scientist with the VTT Technical Research Centre of Finland, Espoo, Finland. Her research interest includes power system communication and automation, 5G and beyond for critical data exchange, distributed control, and real-time systems. 
\end{IEEEbiographynophoto}
\begin{IEEEbiographynophoto}
{Diomidis Michalopoulos}[S'05, M'10, SM'15] (diomidis.michalopoulos@nokia-bell-labs.com) is Department Head of Device Standardization Research, Nokia, Germany. He and his team conduct research on 5GAdvanced/6G networks and devices, with emphasis on physical layer and radio access aspects. Prior to joining Nokia he was employed by the University of British Columbia, Canada, and the University of Erlangen-Nuremberg, Germany. Diomidis obtained the Engineering and PhD degree from the Aristotle University of Thessaloniki, Greece. He received the Marconi Young Scholar award from the Marconi Society and various prizes for academic excellence, including the Banting fellowship in Canada. Diomidis is currently the industry-academia collaboration coordinator within the IEEE EMEA region.
\end{IEEEbiographynophoto}
\begin{IEEEbiographynophoto}{Dem\'ostenes Z. Rodr\'iguez} 
(M'12-SM'15) received the B.S. degree in electronic engineering from the Pontifical Catholic University of Peru, and his M.Sc. and Ph.D. degree from the University of São Paulo in 2009 and 2013, respectively. He is currently an Adjunct Professor with the Department of Computer Science, Federal University of Lavras, Brazil. He has a solid knowledge in Telecommunication Systems based on 15 years of professional experience. His research interest includes QoS-QoE in multimedia services and new generation networks. 
\end{IEEEbiographynophoto} 
\begin{IEEEbiographynophoto}{Heli Kokkoniemi-Tarkkanen}
(heli.kokkoniemi-tarkkanen@vtt.fi) received the M.Sc. in applied mathematics and computer science from the University of Jyväskylä, Jyväskylä, Finland in 1995. Since 1992, she has been working at VTT Technical Research Centre of Finland in several positions, currently as a Senior Scientist. She has over 28 years of experience in commercial, military, and research projects covering various aspects of wireless communication from radio wave propagation modeling and network simulation to early-phase product development. In recent years, she has been focusing on QoS, latency, and reliability aspects by piloting and testing 5G services in new mission-critical vertical use cases such as protection and control of smart energy grids and harbor automation.
\end{IEEEbiographynophoto}
\begin{IEEEbiographynophoto}{Kimmo Ahola}
(kimmo.ahola@vtt.fi) received the M.Sc. degree from the University of Jyväskylä, Jyväskylä, Finland in 1997. From 1996 to 1997, he was a Research Trainee with VTT Technical Research Centre of Finland. From 1997 to 2001, he worked as a Research Scientist with VTT Technical Research Centre of Finland. Since 2001, he has been working as a Senior Scientist and between 2006 and 2013 as a Team Leader in Adaptive Networks team. He has participated and led software development in several national and European projects. Lately, his research interests have focused on 5G networks, software defined networking (SDN), network functions virtualization (NFV), cloud infrastructures, and security in network and cloud infrastructures.
\end{IEEEbiographynophoto}
\begin{IEEEbiographynophoto}{Pedro H. J. Nardelli}[M'07, SM'19]
(pedro.nardelli@lut.fi) is Associate Professor (tenure-track) at LUT University and also Academy of Finland Research Fellow. He is also the coordinator of the Strategic Research Area for Energy Vertical of the 6G Flagship at University of Oulu. 
\end{IEEEbiographynophoto}
\begin{IEEEbiographynophoto}{Gustavo Fraidenraich}
(gf@decom.fee.unicamp.br) graduated in Electrical Engineering from the Federal University of Pernambuco, UFPE, Brazil, in 1997. He received his M.Sc. and Ph.D. degrees from the State University of Campinas, UNICAMP, Brazil, in 2002 and 2006, respectively. From 2006 to 2008, he worked as a Postdoctoral Fellow at Stanford University (Star Lab Group) - USA. Currently, Dr. Fraidenraich is Assistant Professor at UNICAMP - Brazil and his research interests include Multiple Antenna Systems, Cooperative systems, Radar Systems and Wireless Communications in general. He has been associated editor of the ETT journal for many years. Dr. Fraidenraich was a recipient of the FAPESP (Fundação de Amparo \`a Pesquisa do Estado de S\~ao Paulo) young researcher Scholarship in 2009.
\end{IEEEbiographynophoto}
\begin{IEEEbiographynophoto}{Petar Popovski}[S'97, A'98, M'04, SM'10, F'16]
(petarp@es.aau.dk) is a Professor at Aalborg University, where he heads the section on Connectivity and a Visiting Excellence Chair at the University of Bremen. He received his Dipl.-Ing and M. Sc. degrees in communication engineering from the University of Sts. Cyril and Methodius in Skopje and the Ph.D. degree from Aalborg University in 2005. He is currently an Editor-in-Chief of IEEEE JOURNAL ON SELECTED AREAS IN COMMUNICATIONS. His research interests are in the area of wireless communication and communication theory. He authored the book ``Wireless Connectivity: An Intuitive and Fundamental Guide'', published by Wiley in 2020.
\end{IEEEbiographynophoto}
\end{document}